# Point cloud-based registration and image fusion between cardiac SPECT MPI and CTA

TANG Shaojie[1,2], MIAO Penpen[1], GAO Xingyu[1], ZHONG Yu[3], ZHU Dantong[1], WEN Haixing[1], XU Zhihui[4], WEI Qiuyue[1,2], YAO Hongping[5], HUANG Xin[6], GAO Rui[7], ZHAO Chen[8], ZHOU Weihua[9]

1. School of Automation, Xi'an University of Posts and Telecommunications, Xi'an, Shaanxi, 710121, China;
2. Xi'an Key Laboratory of Advanced Control and Intelligent Process, Xi'an, Shaanxi, 710121, China;
3. Second Clinical Medical School of Nanjing Medical University, Nanjing, Jiangsu, 210011, China;
4. Department of Cardiology, The First Affiliated Hospital of Nanjing Medical University, Nanjing, Jiangsu, 210000, China;
5. Department of Pharmacy, The First Affiliated Hospital of Xi'an Jiaotong University, Xi'an, Shaanxi 710061, China;
6. Department of Cardiology, The First Affiliated Hospital of Xi'an Jiaotong University, Xi'an, Shaanxi 710061, China;
7. Department of Nuclear Medicine, The First Affiliated Hospital of Xi'an Jiaotong University, Xi'an, Shaanxi, 710061, China;
8. Department of Computer Science, Kennesaw State University, Marietta, GA, 30060, USA;
9. Department of Applied Computing, Center for Biocomputing and Digital Health, Institute of Computing and Cybersystems, and Health Research Institute, Michigan Technological University, Houghton, MI, 49931, USA

*Corresponding Authors:
Shaojie Tang, PhD                        E-Mail: tangshaojie@xupt.edu.cn
School of Automation, Xi'an University of Posts and Telecommunications
Xi'an, Shaanxi, 710121, China

Zhihui Xu, MD                            E-mail: wx_xzh@njmu.edu.cn
Department of Cardiology, The First Affiliated Hospital of Nanjing Medical University,
300 Guangzhou Rd, Gulou, Nanjing, 210000, China;
Tel: (+86)02568303120

**Abstract:** A method was proposed for the point cloud-based registration and image fusion between cardiac single photon emission computed tomography (SPECT) myocardial perfusion images (MPI) and cardiac computed tomography angiograms (CTA). Firstly, the left ventricle (LV) epicardial regions (LVERs) in SPECT and CTA images were segmented by using different U-Net neural networks trained to generate the point clouds of the LV epicardial contours (LVECs). Secondly, according to the characteristics of cardiac anatomy, the special points of anterior and posterior interventricular grooves (APIGs) were manually marked in both SPECT and CTA image volumes. Thirdly, we developed an in-house program for coarsely registering the special points of APIGs to ensure a correct cardiac orientation alignment between SPECT and CTA images. Fourthly, we employed ICP, SICP or CPD algorithm to achieve a fine registration for the point clouds (together with the special points of APIGs) of the LV epicardial surfaces (LVERs) in SPECT and CTA images. Finally, the image fusion between SPECT and CTA was realized after the fine registration. The experimental results showed that the cardiac orientation was aligned well and the mean distance error of the optimal registration method (CPD with affine transform) was consistently less than 3 mm. The proposed method could effectively fuse the structures from cardiac CTA and SPECT functional images, and demonstrated a potential in assisting in accurate diagnosis of cardiac diseases by combining complementary advantages of the two imaging modalities.
**Keywords：** CTA; SPECT; MPI; point cloud; image registration; image fusion

## 1. Introduction

With the development of social economy and the acceleration of population aging and urbanization, the unhealthy lifestyle of residents has become increasingly prominent, and the incidence of cardiovascular diseases continues to increase [1]. At present, cardiovascular disease is the first cause of death among urban and rural residents in China [2]. The burden of cardiovascular diseases on residents and society is increasing. As different imaging modalities, single photon emission computed tomography (SPECT) myocardial perfusion imaging (MPI) and computed tomography (CT) angiograms (CTA) are widely used in the detection and diagnosis of cardiac diseases [3]. SPECT uses single photon radionuclides injected into human body to emit γ-ray passing through human body, can well reflect the life metabolism of human organs and tissues, and realize the quantitative assessment of myocardial activity and blood perfusion. It is called functional imaging, but its image resolution is low. CTA uses intravenous injection of contrast agent, and then employs CT technique to generate images with high resolution, which can clearly reflect the cardiac structural information, and so is called structural imaging. Generally, CTA is performed independently of SPECT, so the image data are not inherently registered, which is extremely inconvenient for physicians to diagnose cardiac diseases accurately. It is easy to understand that, if the SPECT functional and CTA structural images can be accurately registered and effectively fused together, the advantages of the two imaging modalities can be combined, which is definitely useful for the accurate diagnosis of cardiac diseases [4].

This work developed a coarse-to-fine registration and fusion method for cardiac SPECT and CTA images based on their point clouds. Different U-Net neural networks [5] were trained and used to segment the LV epicardial regions (LVERs) in SPECT and CTA images to generate the point clouds of the LV epicardial contours (LVECs) corresponding to LVERs. Then, the special points of anterior and posterior interventricular grooves (APIGs) were manually marked in both SPECT and CTA images. A coarse registration of special points of APIGs was carried out to ensure a correct cardiac orientation alignment between SPECT and CTA images, and a fine registration by using Iterative Closest Point (ICP) [9], Scale Iterative Closest Point (SICP) [10] or Coherent Point Drift (CPD) [11] algorithm was exploited for the point clouds (together with the special points of APIGs) of the LV epicardial surfaces (LVES, i.e., all LVECs) in CTA and SPECT images. Finally, the original three-dimensional (3D) CTA and SPECT images were registered by exploiting the spatial transform matrix acquired by the point cloud registration, and SPECT image data were rendered to the CTA's whole heart epicardial surface (WHES) segmented by the region growth algorithm [7] to achieve the expected registration and fusion of SPECT and CTA images.

## 2. Methodology

### 2.1 Workflow

Firstly, the LV and whole heart in CTA images were segmented to generate the LVERs and whole heart epicardial region (WHER), and their respective point clouds of LVECs and WHES, respectively. Secondly, after the LVERs were segmented in the cardiac SPECT images, the point cloud of the corresponding LVECs was generated. Then, the point cloud of the LVES (i.e., all LVECs) of CTA was registered with that of SPECT by using a point cloud registration algorithm. Finally, the spatial transform acquired by point cloud registration was directly exploited to achieve a multimodal image registration and fusion between the original 3D CTA and SPECT images. At the same time, SPECT myocardial perfusion imaging (MPI) data could be mapped onto the CTA's WHES to achieve the expected fusion of SPECT functional data and CTA structural information. The workflow of the proposed method is shown in Fig. 1.

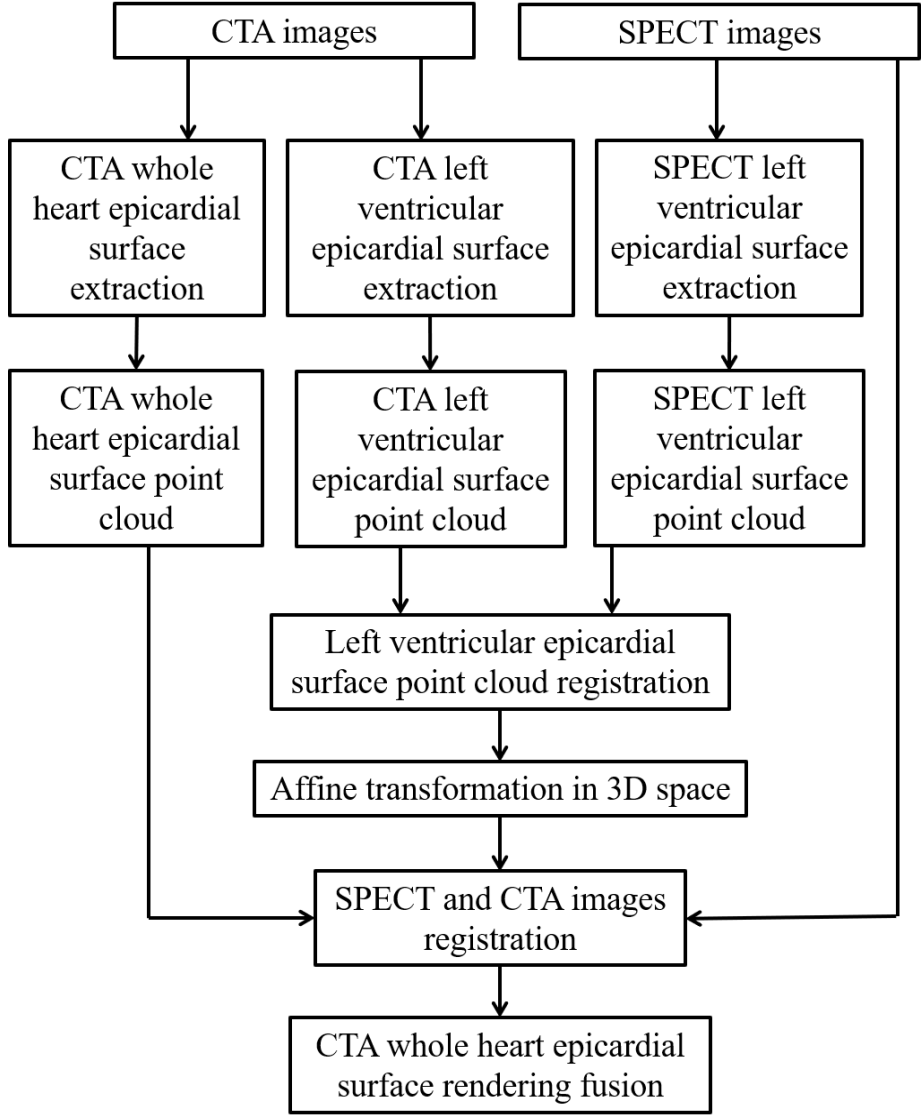

Fig. 1 Workflow of the proposed method

**2.2 Point clouds extraction**

**2.2.1 CTA left ventricular epicardial surface extraction**

A U-Net neural network was implemented to segment the LVERs using a Pytorch framework. The trunk feature extraction part of U-Net was similar to that of VGG, so VGG16 [21] was used as the backbone to facilitate the use of its pre-training weight. In the enhanced feature extraction part, the feature layer obtained after up-sampling had the same width and height as the input image. 1×1 convolution was used for channel adjustment, and the number of channels in the final feature layer was adjusted into the class number. In this work, the image regions were classified into LV and background, so the class number was 2. The structure of the adopted U-Net neural network is shown in Fig. 2.

To acquire the ground truth for training the segmentation model, the segmentation module of

3D Slicer software [6] was used to manually segment the LVER in CTA images. A total of 69 patients' data and 16,896 CTA images were manually segmented. All 14,250 images from 58 patients' data were randomly divided into training set and validation set according to 9:1 ratio. The test set included 2,646 images from 11 patients' data. The input and label images were both in a size of 512 × 512 pixels.

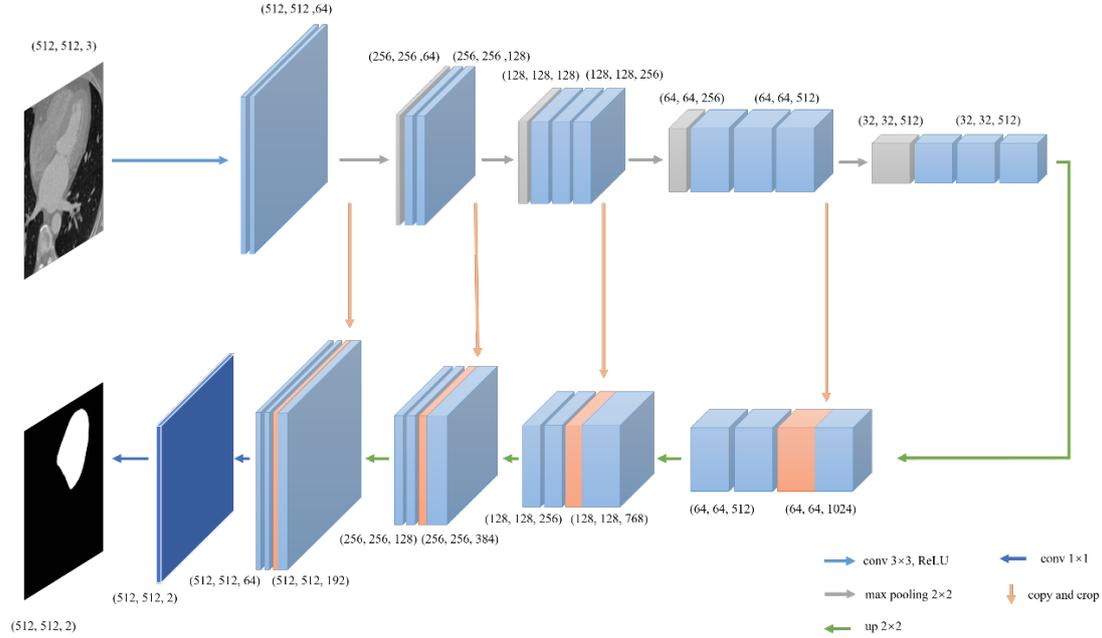

Fig. 2 Structure of U-Net neural networks

With the trained model, the segmentation result was predicted on the test set. The average similarity between the segmentation results on the test set and the ground truth was calculated with DSC defined as follows,

$$DSC = \frac{2|A \cap B|}{|A| + |B|} \tag{1}$$

After calculation, the average similarity on the test set is 0.9565 +/- 0.1523. The segmentation results were then used to generate the point cloud of CTA LVECs with the original CTA image coordinates, saved as a file with .stl format, and displayed with VTK rendering in 3D space.

Furthermore, an in-house program was developed for manually marking the special points of the APIGs in the CTA images and integrating them into the point cloud of CTA LVECs, as shown in Fig. 3.

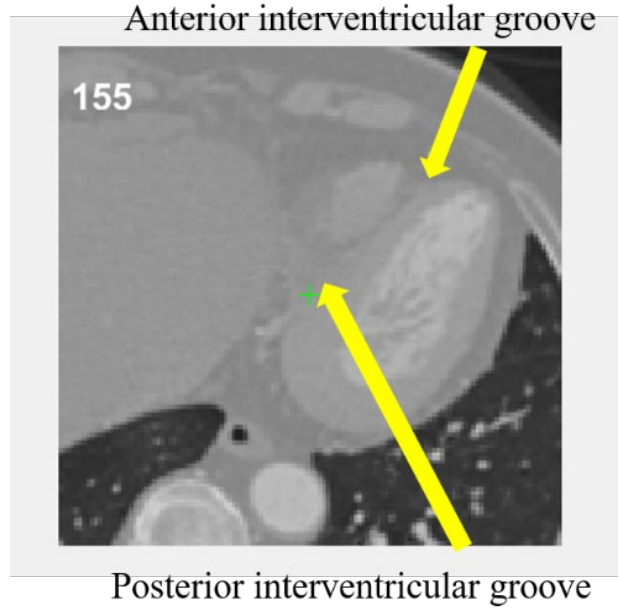
Fig. 3 Special points of anterior and posterior interventricular grooves in a CTA image

**2.2.2 CTA whole heart epicardial surface extraction**

Region growth algorithm [12] was used for image region segmentation. The basic idea is to assemble pixels with similar properties to form regions. This algorithm requires to select the initial growth points (i.e., seed points) in the regions to be segmented, and then combine the seed points and the pixels in the neighbor that meet the growth rule into the regions where the seed points are located. As there are no more pixels that meet the growth rule, the algorithm ends.

An in-house program was developed for segmenting the whole heart epicardial region (WHER) in CTA images with the region growth algorithm, as shown in Fig. 4. Based on the gray level difference criterion [7], let's set the average gray level value of the segmented pixels as $M$, and the gray level value of the current pixel as $v$. If $|v-M| \leq T$, the current pixel will be merged into the region and $M$ is updated to

$$M = (n \times M + v)/(n+1) \qquad (2)$$

where $T$ is the threshold value and $n$ is the number of the pixels that have been merged into the region; otherwise, the current pixel will not be merged into the region. As a rule of thumb, the threshold value $T$ was set to 400. Isosurfaces were extracted from the segmented regions, the STL model composed of triangular patches and vertices was returned, and saved as a file with .stl format too.

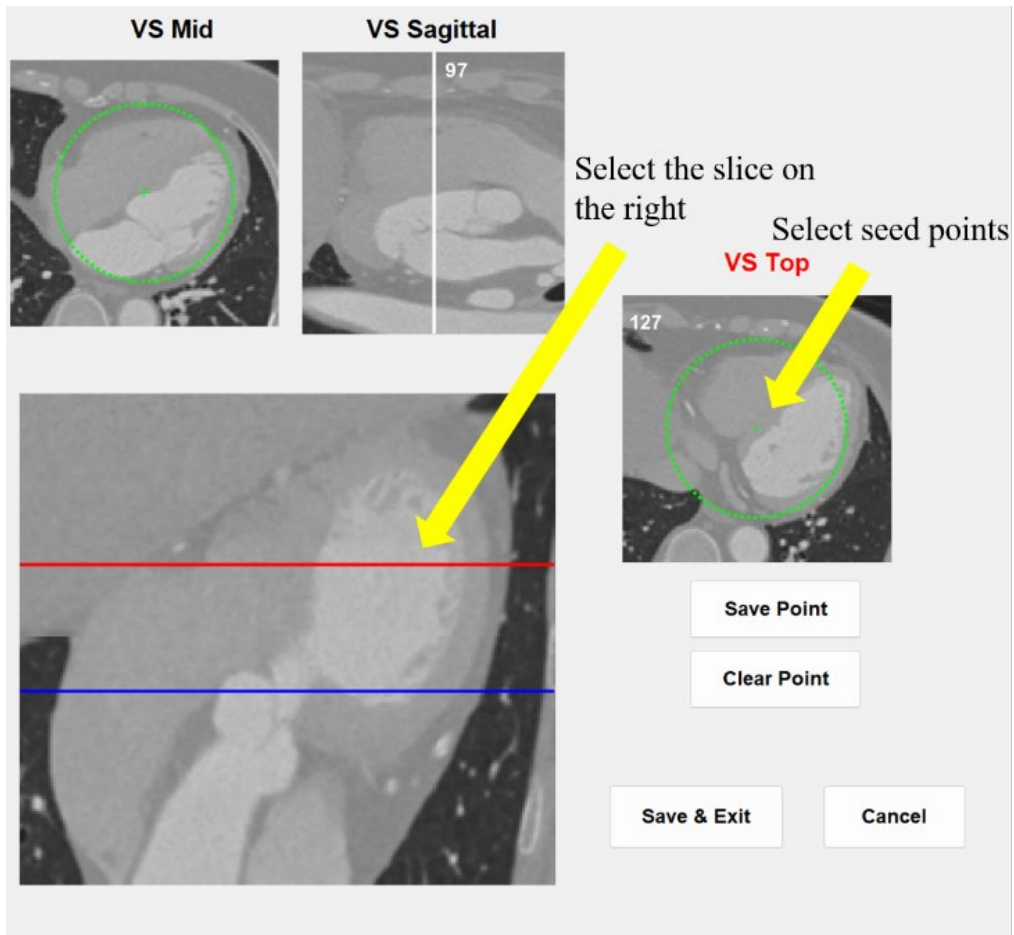

Fig. 4 Whole heart epicardial region was segmented in CTA images with the region growth algorithm

**2.2.3 SPECT left ventricular epicardial surface extraction**

An in-house program was developed to set LV positions in SPECT images [13], including LV's center, apex, base, etc., as shown in Fig. 5. After the positions were set, the SPECT images were converted from the original short-axis form to the long-axis from. That is, the planes, whose intersection was the line formed from the LV's center to the LV's apex, were sampled every 9° along the longitude direction, and so there were totally 20 planes per LV [14]. Each plane was converted from Cartesian coordinates to polar coordinates. Then, another in-house program was used to automatically segment the LVERs. All 5,600 images from 35 patients were acquired as the training set. In the Python3, the LVERs were segmented using another U-Net neural networks trained [15]. Its input layer was of 64 × 64 pixels. The obtained LVERs was converted back from polar coordinates to Cartesian coordinates, ensuring that the coordinates of the generated point cloud of the LVECs were consistent with those in the original SPECT images.

On the basis of the in-house program for setting LV positions, the function of manual point selection was added as shown in Fig. 6. After selecting the special points of the APIGs, we integrated them into the point cloud of the LVECs.

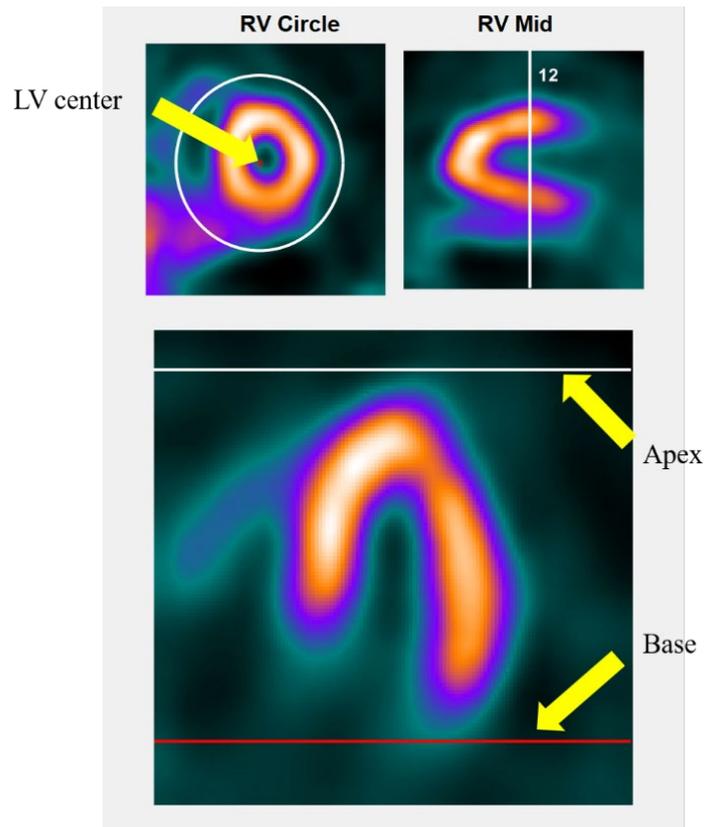

Fig. 5 Setting left ventricular positions in a SPECT image

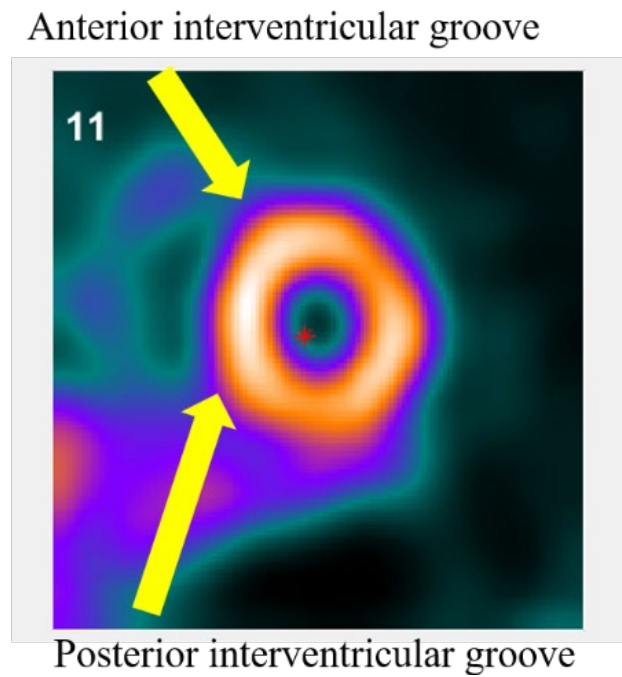

Fig. 6 Special points of anterior and posterior interventricular grooves in a SPECT image

## 2.3 Point cloud registration
### 2.3.1 Coarse registration

Since LV has no obvious features in shape, a direct use of point cloud registration [16] will generally lead to a wrong registration, as shown in Fig. 7. It can be observed that, the point clouds of LVECs in SPECT and CTA do not match well, resulting in a registration error.

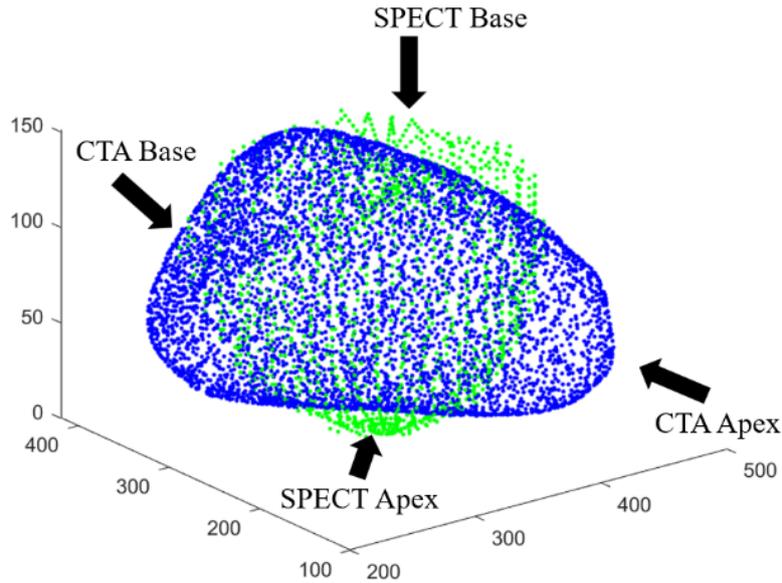

Fig. 7 An instance of wrong registration

According to the anatomical characteristics of heart, the anterior interventricular groove [17] is a shallow one extending from the apex of the coronal groove on the thoracic costal surface of the heart, consistent with the front edge of the ventricular septum, and the boundary between the LV and right ventricle (RV) on the anterior surface of the heart. The posterior interventricular groove is a shallow one extending from the coronal groove to the apex of the heart on the diaphragmatic surface. It is consistent with the lower margin of the ventricular septum and is the boundary between the LV and RV on the inferior surface of the heart. Therefore, the special points of APIGs in SPECT and CTA images can be used to ensure the correct alignment of the left and right sides, the apex and the base of the heart during data preprocessing.

So, a coarse procedure was developed for registering the special points of APIGs in SPECT and CTA images. Firstly, we intentionally saved the special points following a storage order from AIG to PIG, as shown in Fig. 8. Then, the special points of APIGs in SPECT and CTA were down-sampled into the same number $m$ and denoted respectively as: $p_i(i=1,2,\cdots,m)$ and $c_i(i=1,2,\cdots,m)$. The spatial transform matrix containing rotation, translation and scaling could be determined by using a Umeyama algorithm [22] in Open3D [8]. The transform matrix determined was then applied as an initial to the fine registration of the point clouds of LVECs of SPECT and CTA.

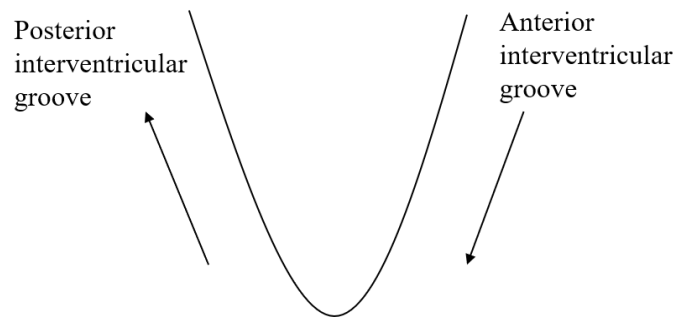

Fig. 8 Storage order of the special points in anterior and posterior interventricular grooves in both SPECT and CTA images

**2.3.2 Fine registration**

On the basis of coarse registration, ICP [9], SICP [10] or CPD [11] algorithm could be used for the fine registration.

ICP is an iterative rigid registration algorithm [9]. The core idea of the algorithm is to make the corresponding point pairs in the two point clouds as close as possible in space, that is, to minimize the following objective function:

$$f(\boldsymbol{R},\boldsymbol{T}) = \sum_{i=1}^{N} \left\| p_t^i - \boldsymbol{R} \cdot p_s^i - \boldsymbol{t} \right\|^2 \quad (3)$$

where $\boldsymbol{R}$ represents the rotation matrix, $\boldsymbol{t}$ represents the translation vector, $p_s^i$ is the i-th point in the source point cloud $p_s$, $p_t^i$ is the i-th point in the target point cloud $p_t$, $N$ represents the number of corresponding point pairs in the original and the target point clouds. The algorithm needs to initialize the rotation matrix $\boldsymbol{R}$ and translation vector $\boldsymbol{t}$, and then calculate the Euclidean distance error between the original and the target point clouds after the rotation and translation transformations. The rotation matrix $\boldsymbol{R}$ and translation vector $\boldsymbol{t}$ are optimized by Singular Value Decomposition (SVD) and re-iterated until the calculated Euclidean distance error meets the stopping requirements. The iteration is stopped to obtain the finally optimal rotation matrix $\boldsymbol{R}$ and translation vector $\boldsymbol{t}$. Since ICP algorithm can only satisfy the registration between two point clouds with the same scale, SICP algorithm was proposed for point cloud registration with different scales [10].

SICP algorithm [10] introduces a scale factor on the basis of ICP algorithm, which can realize point cloud registration with different scales, and form optimization problems about the scale factor $s$, rotation matrix $\boldsymbol{R}$ and translation vector $\boldsymbol{t}$. The objective function after adding the scaling factor $s$ is defined as:

$$f(\boldsymbol{R},\boldsymbol{T},s) = \sum_{i=1}^{N} \left\| s \cdot \boldsymbol{R} \cdot p_s^i + \boldsymbol{t} - p_t^i \right\|^2 \quad (4)$$

The scaling factor $s$ in the original SICP algorithm is isotropic, that is, the scaling ratios in the $x$, $y$ and $z$ directions are the same. The scale factor $s$ is modified into its anisotropy counterpart (i.e., a 3D vector ($s_1$, $s_2$, $s_3$)), so that independent scaling in $x$, $y$ and $z$ directions can be realized. The corresponding objective function is updated as:

$$f(\boldsymbol{R},\boldsymbol{T},s) = \sum_{i=1}^{N} \left\| \begin{pmatrix} s_1 & 0 & 0 \\ 0 & s_2 & 0 \\ 0 & 0 & s_3 \end{pmatrix} \cdot \boldsymbol{R} \cdot p_s^i + \boldsymbol{t} - p_t^i \right\|^2 \quad (5)$$

The distance between SPECT and CTA point clouds is minimized by still using SVD method to solve the related parameters. When the preset threshold is reached, the iteration stops and the transformation parameters are obtained.

CPD is a point cloud registration algorithm based on the Gaussian Mixed Model (GMM) [11]. In this algorithm, the registration of two point clouds is regarded as a probability density estimation problem. One point cloud represents the GMM centroid, namely the source point cloud. The other represents data points, namely the target point cloud. The source point cloud is fitted to the target point cloud by maximizing the likelihood function and the Expectation-Maximization (EM) algorithm is used to solve the transformation matrix. The core idea of this algorithm is to define the correspondence between two clouds as a posteriori probability, and express "distance" with "probability". The closer the distance, the greater the contribution to probability. CPD algorithm can constrain a registration with rigid, affine and non-rigid transforms. The rigid transform is defined as:

$$T(\boldsymbol{R},\boldsymbol{t},s) = s \cdot \boldsymbol{R} \cdot p_s + \boldsymbol{t} \qquad (6)$$

The affine transform is defined as:

$$T(\boldsymbol{B},\boldsymbol{t}) = \boldsymbol{B} \cdot p_s + \boldsymbol{t} \qquad (7)$$

where $\boldsymbol{B}$ is the affine transform matrix. The CPD with rigid and affine transforms were used in this work.

**2.4 3D image registration and fusion**

During the preprocessing, the coordinates of SPECT and CTA point clouds were given in the coordinate systems of their original images, respectively, so the transform matrix obtained after point cloud registration can also be applied to the registration of their original 3D images. The imwarp function [19] in MATLAB can apply geometric transform to the image. The imwarp function's input is the spatial reference image and its associated spatial reference object specified by the geometric transform matrix and the image data; the imwarp function's output is the space reference image specified by the transformed image data and its associated space reference object, combined with 'OutputView' which can limit the size and position of the output image in the world coordinate system. The spatial reference object defines the position of the image in the world coordinate system, which is constructed by the parameters listed in Table 1, where the abbreviations of the parameters are defined in parentheses. These parameters can be calculated by the voxel size and *(L0, P0, S0)*, that is, the anatomical coordinates of the first voxel (0,0,0):

$$\begin{cases} XWL = [L0, L0 + IWX] \\ YWL = [P0, P0 + IWY] \\ ZWL = [S0, S0 + IWZ] \end{cases} \qquad (8)$$

where $(IWX, IWY, IWZ) = IS \cdot (PWX, PWY, PWZ)$.

Since the point clouds of the LVES and the WHES acquired from CTA image are defined in the same coordinate system, the transform matrix acquired by LV point cloud registration between SPECT LVES and CTA LVES can be directly applied to the point cloud registration between SPECT LVES and CTA WHES, so as to establish their mapping relationship. In addition, the point cloud coordinates of SPECT and CTA during preprocessing are corresponding to their original image coordinates, respectively, so the myocardial perfusion data extracted from the SPECT image can be conveniently fused to the CTA WHES with the help of point cloud registration transform matrix.

Table 1 Spatial reference object parameters

| | |
|---|---|
| ImageSize(IS) | Size of image |
| PixelExtentInWorldX(PWX) | The size of a pixel along the X dimension in the world coordinate system |
| PixelExtentInWorldY(PWY) | The size of a pixel along the Y dimension in the world coordinate system |
| PixelExtentInWorldZ(PWZ) | The size of a pixel along the Z dimension in the world coordinate system |
| XWorldLimits(XWL) | The starting and ending positions of images along the X dimension in the world coordinate system |
| YWorldLimits(YWL) | The starting and ending positions of images along the Y dimension in the world coordinate system |
| ZWorldLimits(ZWL) | The starting and ending positions of images along the Z dimension in the world coordinate system |
| ImageExtentInWorldX(IWX) | The size of the image in the world coordinate system along the X dimension |
| ImageExtentInWorldY(IWY) | The size of the image in the world coordinate system along the Y dimension |
| ImageExtentInWorldZ(IWZ) | The size of the image in the world coordinate system along the Z dimension |

## 3 Experimental results and analysis
### 3.1 Experimental process

The experiment was carried out in the programming environment of MATLAB2020a and Python3. The complete process was shown below using data from one patient as an instance. In this patient, the SPECT image volume was of 64 × 64 × 27 voxels, with each a size of 6.4 × 6.4 × 6.4 mm$^3$; while CTA image volume was of 512 × 512 × 253 voxels, with each a size of 0.4082 × 0.4082 × 0.5mm$^3$. (L0, P0, S0) was (-81.1, -278.1, -234).

The curves of loss values in the U-Net neural network training process are shown in Fig. 9. As the epoch is less than 20, loss decreases rapidly. As the epoch is between 20 and 50, loss decreases relatively slowly. As the epoch is greater than 80, loss tends to be stable. The final loss value is convergent to ∼0.01. The experimental result shows that the model is convergent.

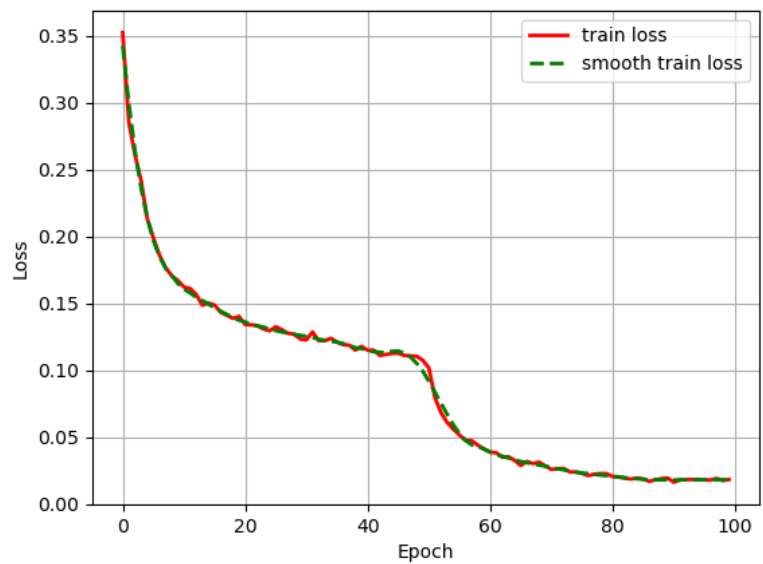

Fig. 9 Loss curves of the U-Net neural network models during training

The U-Net neural network was used to segment the LV in the CTA image, and the point cloud was generated after the output STL model was combined with VTK plane rendering. The special points of APIGs were calibrated and merged into the LV point cloud, as shown in Fig. 10. In the figure, the special points of APIGs were set as black and red, respectively, while the other points were set as blue.

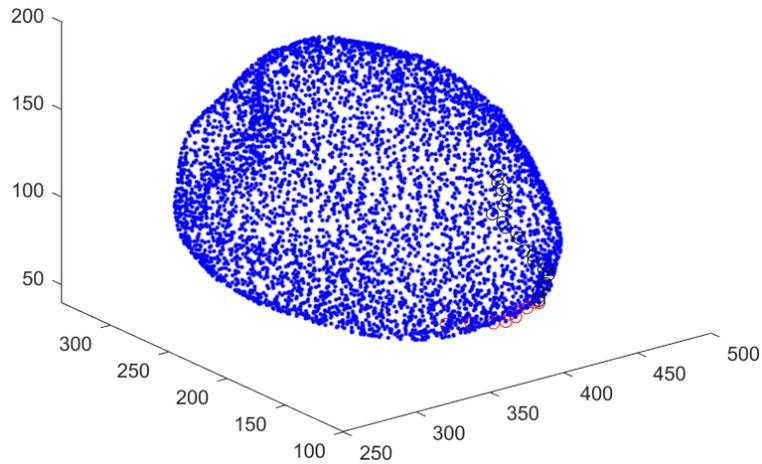

Fig. 10 Point cloud of left ventricular epicardial surface in CTA

Region growth algorithm was used for the extraction of CTA's WHES, and the results were shown in Fig. 11.

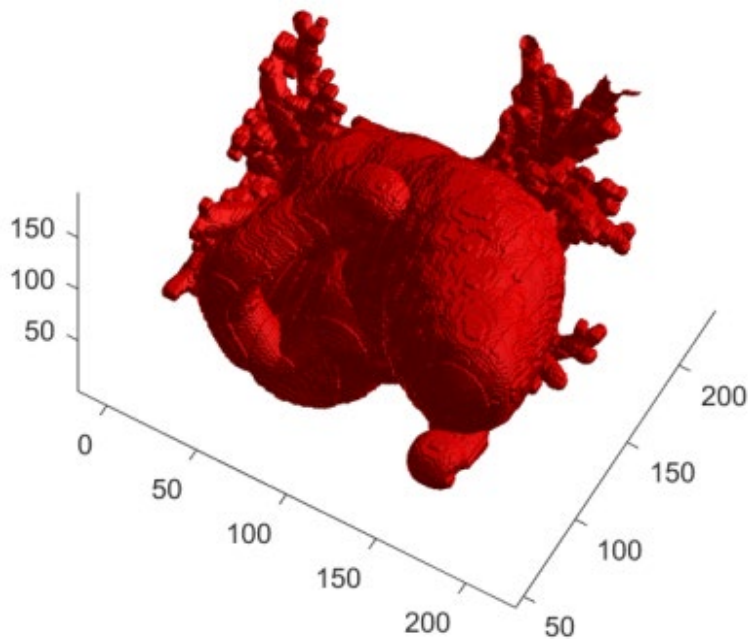

Fig. 11 Whole heart epicadial surface of CTA

U-Net neural network was used to segment the LVERs of SPECT and generate the point cloud of LVECs. The special points of APIGs were marked and merged into the point cloud of LVES (i.e., all LVECs), as shown in Fig. 12.

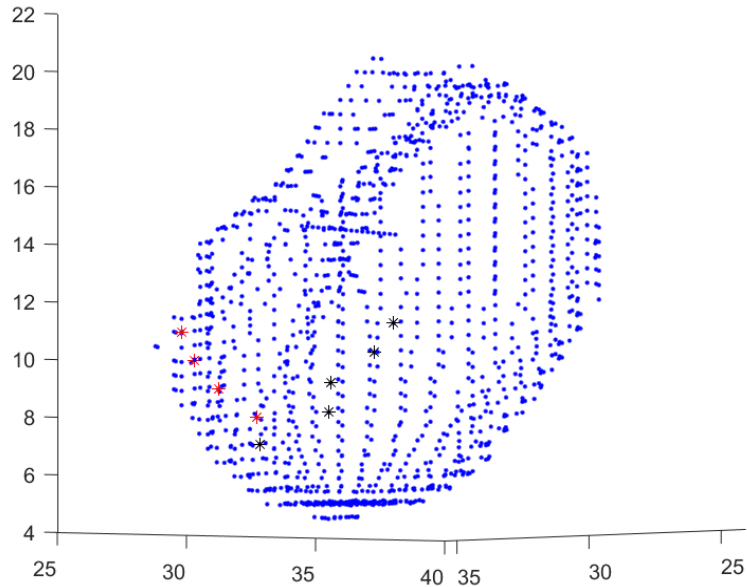

Fig. 12 Point cloud of left ventricular epicardial surface in SPECT

The cases before and after coarse registration based on APIGs were shown in Fig. 13. In the figure, SPECT point cloud was set as green, wherein the special points of AIG were set as black *, while those of PIG were set as red *. The CTA point cloud was set to blue, wherein the special points of AIG were set to black o, while those of PIG were set to red o.

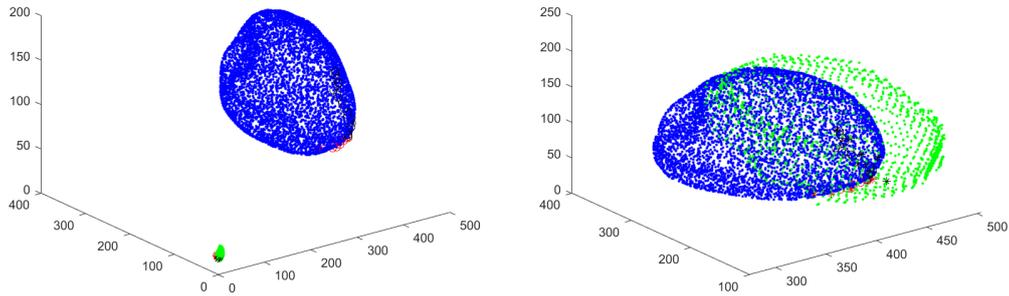

Fig. 13 Results before (left) and after (right) coarse registration

On the basis of coarse registration, ICP, SICP and CPD algorithms were applied, respectively. The fine registration results were shown in Fig. 14. The color and shape of the points in the figure showed the same meanings as in Fig. 13.

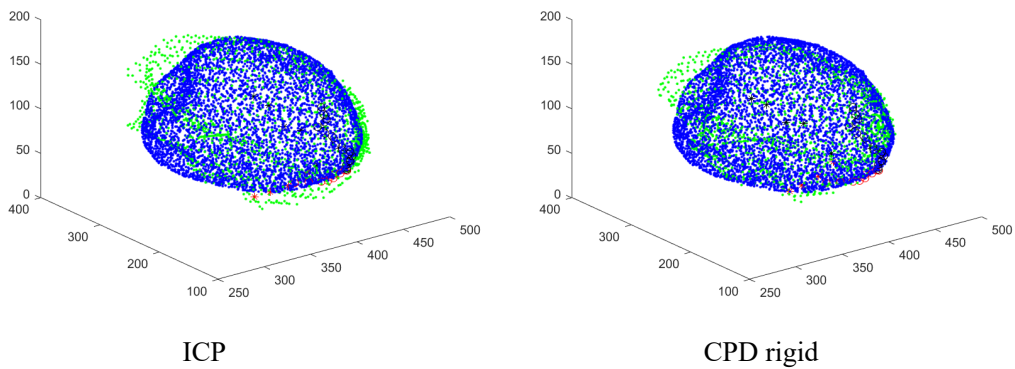

ICP                                              CPD rigid

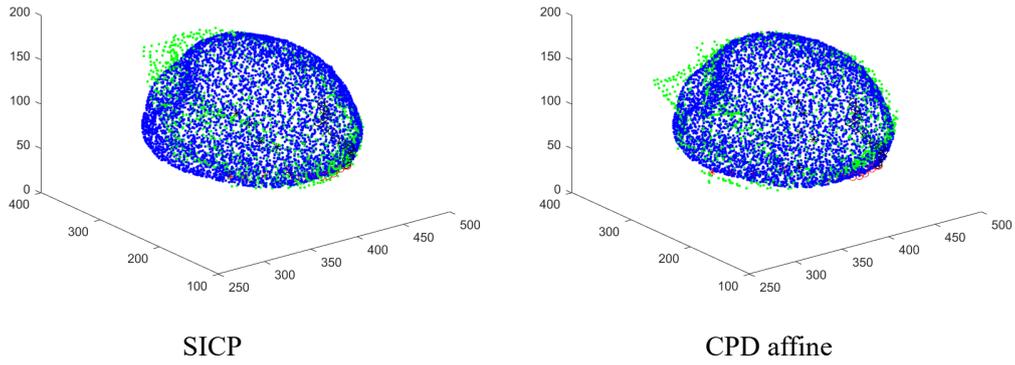

| SICP | CPD affine |

Fig. 14 Fine registration results

According to the image volumes of SPECT and CTA, voxel size and (L0, P0, S0), the spatial reference object parameters were calculated. Then, the spatial reference objects of SPECT and CTA were constructed, and the original 3D SPECT images was transformed and registered by combining the point cloud registration results, as shown in Figs. 15 and 16. Fig. 15 shows the cross-sectional display of image registration results, wherein purple and green colors represent CTA and SPECT images, respectively. Fig. 16 shows the coronal display of image registration results, wherein CTA and SPECT images are displayed on the left and right sides, respectively.

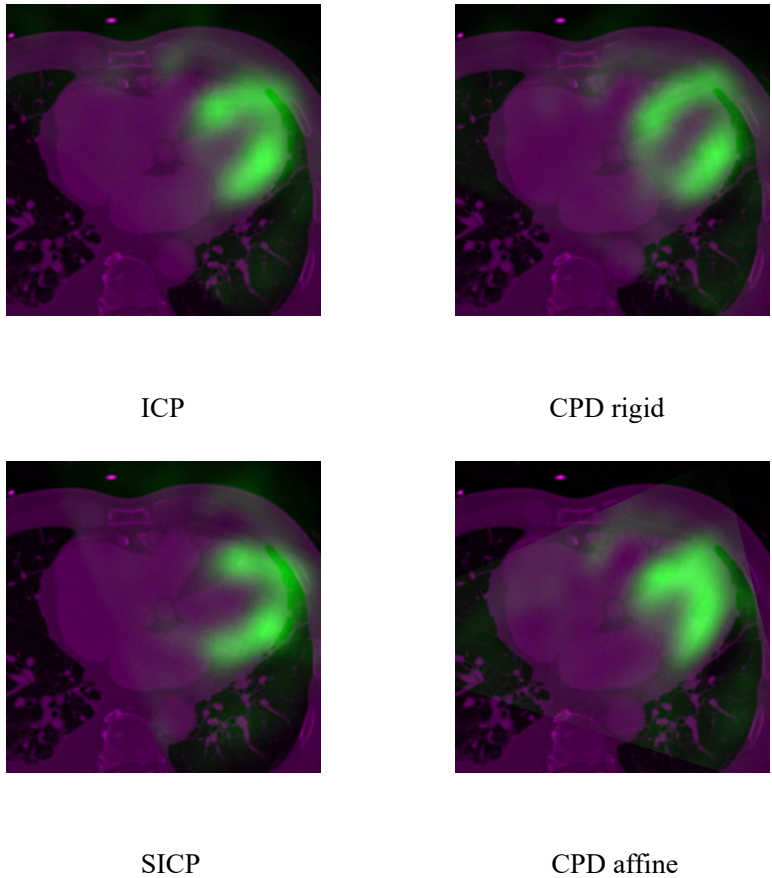

| ICP | CPD rigid |

| SICP | CPD affine |

Fig. 15 Cross-sectional display of registration results

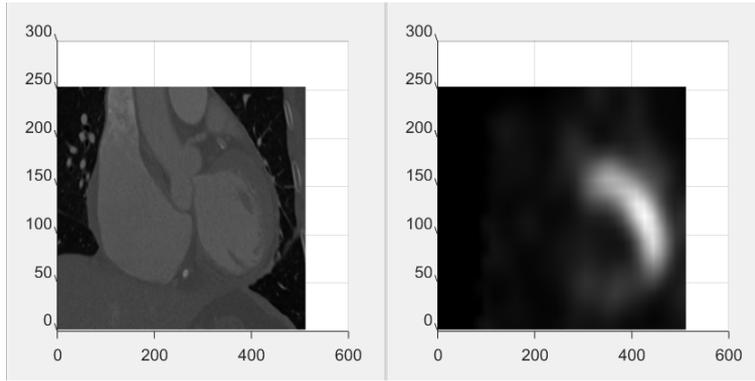
ICP

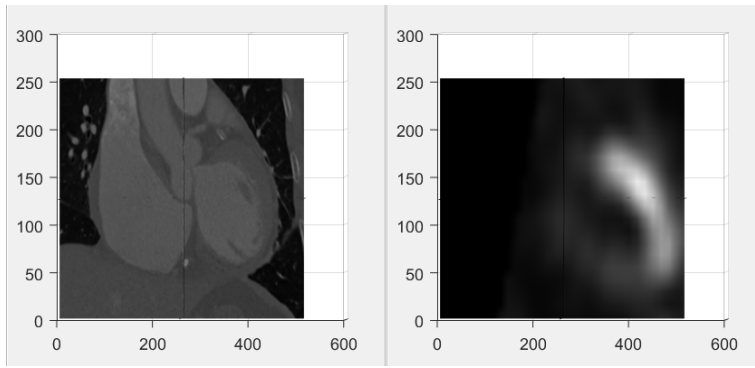
SICP

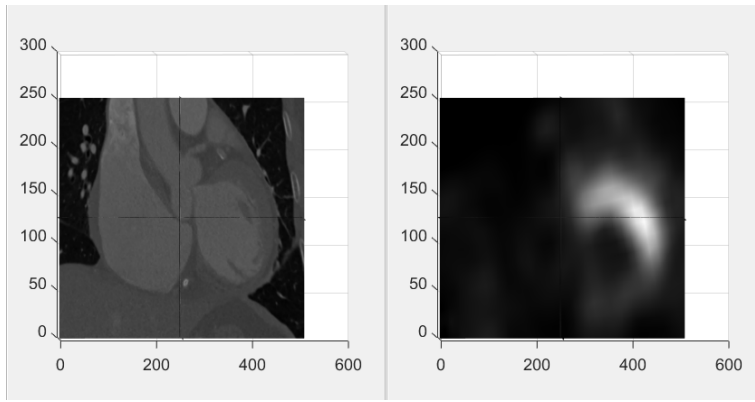
CPD rigid

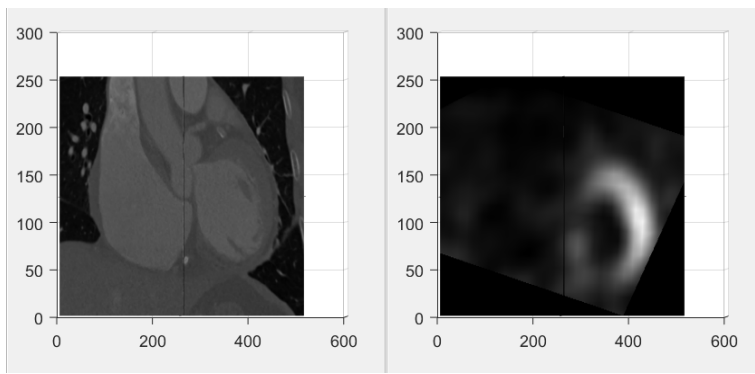
CPD affine

Fig. 16 Coronal plane display of the image registration results

The 3D affine transform acquired was used to map MPI data of SPECT onto the CTA's WHES, and the fusion results were shown in Fig. 17. The brighter the color, the better the cardiac blood flow at that location.

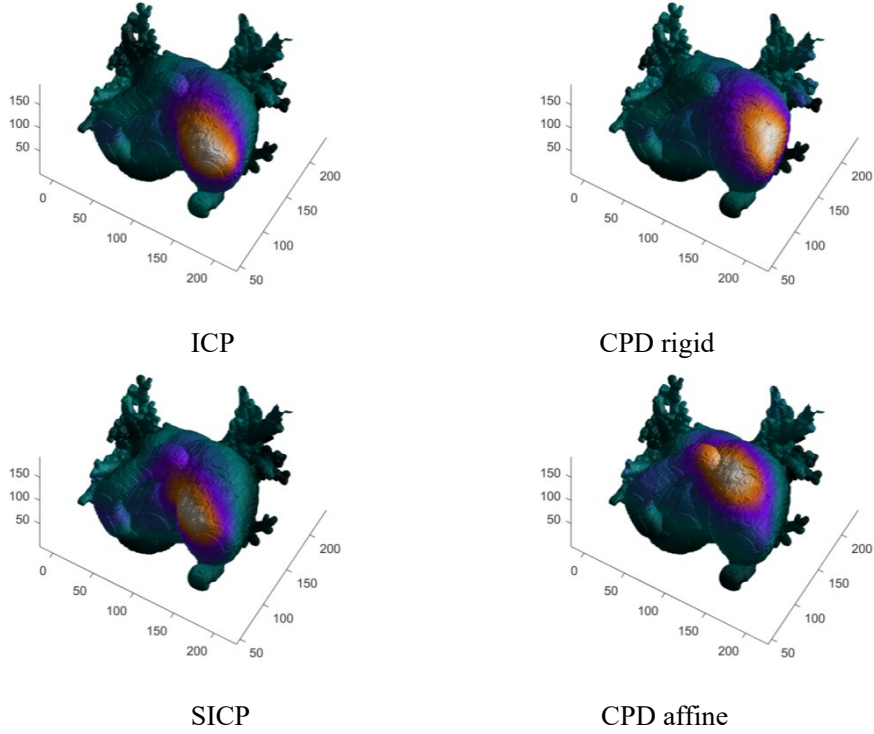

ICP          CPD rigid

SICP          CPD affine

Fig. 17 Fusion results of the SPECT MPI mapped onto the CTA's whole heart epicardial surface

### 3.2 Experimental Analysis

Since the 3D image registration and fusion were realized on the basis of point cloud registration using the obtained transform matrix, the evaluation of the 3D image registration and fusion results was converted to the evaluation of the point cloud registration results. The evaluation criteria were divided into two parts: the first is whether the cardiac orientation is accurately registered. By observing the locations of the heart apex, base, anterior and posterior interventricular grooves, it is demonstrated that the cardiac orientation was accurately registered, and this is a qualitative analysis. The second is quantitative analysis to evaluate the registration accuracy of point clouds, which is defined as the mean distance error between SPECT and CTA point clouds after registration. If the number of SPECT and CTA point clouds is different [19], only the matching points are calculated. The calculation formula is defined as:

$$e = \frac{1}{n}\sum_{i=1}^{n}\|x_i - y_i\| \qquad (9)$$

where $n$ is the number of pairing points, and $x_i$ and $y_i$ represent the spatial coordinates of pairing points in the two point clouds.

As can be seen from Fig. 14 to Fig. 17, after fine registration, the orientations of APIGs are consistent, and the heart registration is more accurate. Fifteen patients were randomly selected for the experiment, and the mean distance error after point cloud registration was calculated. The quantitative results were listed in Fig. 18.

As can be seen from Fig. 18, on the basis of coarse registration, all the four fine registration algorithms can achieve relatively low errors. Meanwhile, the mean distance error of ICP was the largest, whereas that of CPD with affine transform was the smallest. This is because ICP has not considered scaling and can only perform rotation and translation, while CPD with affine transform can be optimized in multiple directions, so the error is reasonably minimal. In general, the registration results can generally ensure that the heart registration is correct with a low mean distance error.

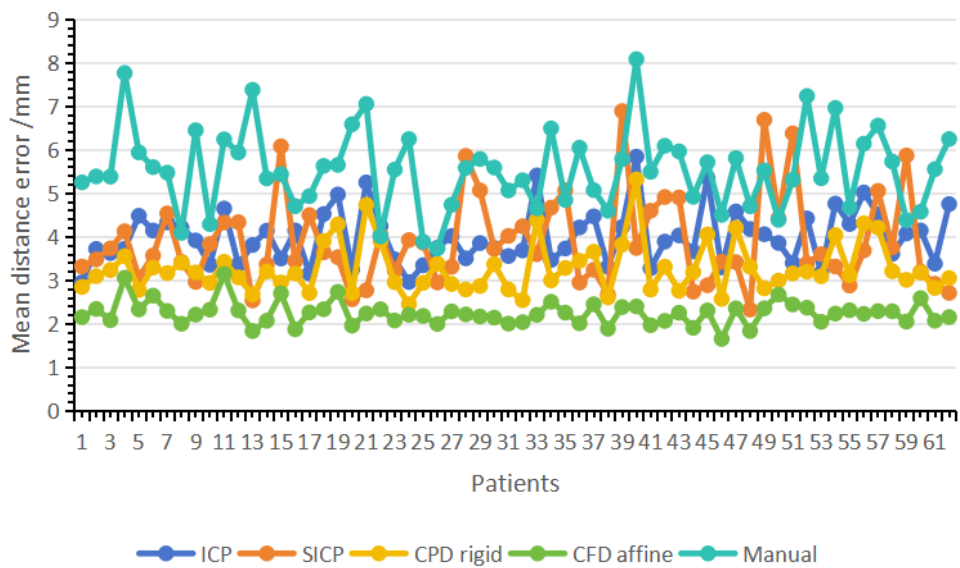

Fig. 18 Mean distance error of point cloud registration

**4 Conclusion**

    In this work, an image information fusion of SPECT MPI and CTA based on point cloud registration was proposed. Firstly, two different U-Net neural networks were used to segment and extract the LVECs in the SPECT and CTA images, respectively, to obtain the SPECT and CTA point clouds. Secondly, an in-house program was developed and used to manually pick up and coarsely register the special points of the APIGs in the SPECT and CTA images. Thirdly, LV point clouds of SPECT and CTA were finely registered based on ICP, SICP and CPD algorithms. Finally, SPECT and CTA images were registered according to the 3D affine transform acquired by the point cloud registration, and the MPI data of SPECT were mapped and fused onto the CTA's WHES. The experimental results showed that the proposed method has higher registration accuracy and better fusion effect while ensuring the correct cardiac orientation in heart registration.